\documentclass{ws-procs9x6}            

\DeclareMathAlphabet{\mathcal}{OMS}{cmsy}{m}{n}
\DeclareMathAlphabet\mathrsfso{U}{rsfso}{m}{n}
\begin{document}
\title{Access to decoupled information \\on Generalized Parton Distributions (GPDs) \\via Double Deeply Virtual Compton Scattering (DDVCS)}

\author{S. Zhao$^*$}

\address{Institut de Physique Nucl\'{e}aire d'Orsay
\\ CNRS/IN2P3, Universit\'{e} Paris Sud \& Paris Saclay
\\ 15 rue Georges Cl\'{e}menceau, 91406 Orsay, France\\
$^*$E-mail: zhao@ipno.in2p3.fr}

\begin{abstract}
The Generalized Parton Distributions (GPDs) constitute an appropriate framework for a universal description of the partonic structure of the nucleon. Double Deeply Virtual Compton Scattering (DDVCS) process provides the only experimental way to measure independently the dependences of the GPDs on the average and transferred momentum. This proceeding discusses the feasibility of a DDVCS experiment in the context of JLab 12 GeV, model-predicted pseudo-data, and extraction of the relevant GPDs information based on a fitter algorithm.
\end{abstract}


\bodymatter

\section{Introduction}

The electroproduction of a lepton pair $eN\rightarrow eNl\bar{l}$, which is sensitive to the DDVCS amplitude, provides the only experimental framework for a decoupled measurement of GPDs($x,\xi,t$) as a function of both the average momentum fraction $x$ and the transferred one $\xi$\cite{DDVCS1,DDVCS2,DDVCS3}. For instance, the Compton form factor (CFF) $\mathcal{H}$ associated with the GPD $H$ and accessible in cross section or beam spin asymmetry experiments can be written
\begin{eqnarray}
\mathcal{H}(\xi',\xi,t)=\mathcal{P}\int_{-1}^1H(x,\xi,t)\bigg[\frac{1}{x-\xi'}+\frac{1}{x+\xi'}\bigg]dx
-i\pi H_+(\xi',\xi,t)
\label{eq1}
\end{eqnarray}
where $\mathcal{P}$ indicates the Cauchy principal value of the integral, and $H_+$ is the singlet GPD defined as $H_+(\xi',\xi,t)=H(\xi',\xi,t)-H(-\xi',\xi,t)$. The real part of the CFF is a complex quantity involving the convolution of parton propagators and the GPD values over $x$, but the imaginary part accesses the GPD values at $x=\pm \xi'$. The famous DVCS process has its limitation of $\xi'=\xi$. Because of the virtuality of final state photons, DDVCS provides a way to circumvent this limitation, allowing access to decoupled information at $\xi'\neq\xi$.

\begin{figure}[b]
\centering
\includegraphics[height=.33\textwidth]{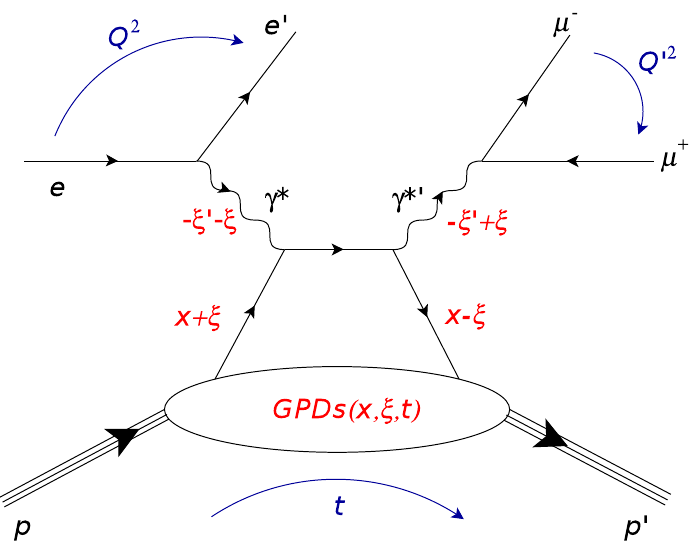}
\includegraphics[height=.33\textwidth]{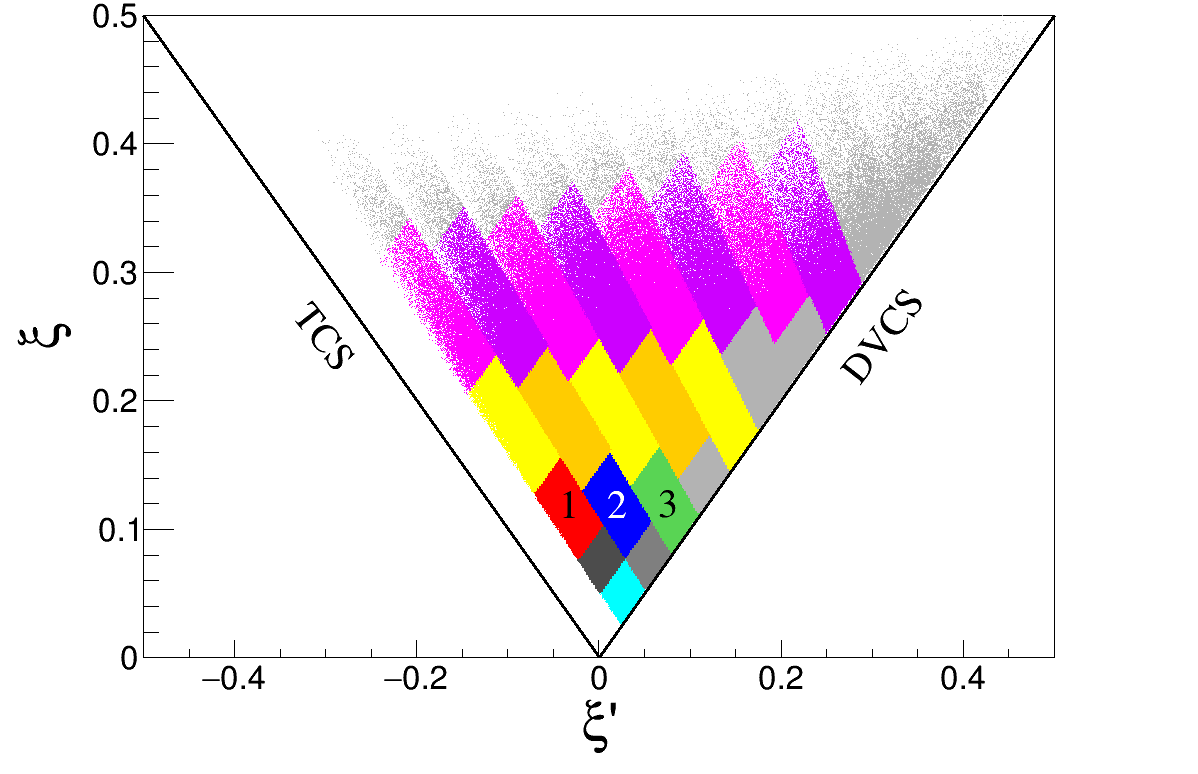}
\caption{\label{fig1} Left panel: the handbag diagram symbolizing the DDVCS direct term with di-muon final states (there is also a crossed diagram where the final time-like photon is emitted from the initial quark). Right panel: The illustration of the grid of bins in ($\xi',\xi$) plane at forward limit ($t=0$), where the solid lines indicate the DVCS ($\xi'=\xi$) and the Time-like Compton Scattering (TCS) limitation ($\xi'=-\xi$). }
\end{figure}

The DDVCS process is very challenging from the experimental point of view due to the small magnitude of the cross section, and requires high luminosity and full exclusivity of the final state. Taking advantage of the energy upgrade of the CEBAF accelerator, it is proposed to investigate the electroproduction of $\mu^+ \mu^-$ di-muon pairs (avoiding complex antisymmetrization issues) and measure the beam spin asymmetry of the exclusive $ep\rightarrow e'p'\gamma^* \rightarrow e'p'\mu^+\mu^-$ reaction in the hard scattering regime \cite{Proposal1,Proposal2,Anikin,ZHAO}.

At sufficiently high virtuality of the initial space-like photon and small enough four-momentum transfer to the nucleon with respect to the photon virtuality ($-t \ll Q^2$), DDVCS can be seen as the absorption of a space-like photon by a parton of the nucleon, followed by the quasi-instantaneous emission of a time-like photon by the same parton, which finally decays into a di-muon pair (left panel of Fig.~\ref{fig1}). $Q^2$ and $Q'^2$ represent the virtuality of the incoming space-like and outgoing time-like photons, respectively. The scaling variable $\xi'$ and $\xi$ write
\begin{eqnarray}
\xi' = \frac{Q^2-Q'^2+t/2}{2Q^2/x_\text{B}-Q^2-Q'^2+t}~\text{and}~
\xi  = \frac{Q^2+Q'^2}{2Q^2/x_\text{B}-Q^2-Q'^2+t}.
\end{eqnarray}

\section{Experiment projections}
\label{sec2}

\begin{figure}[b]
\centering
\includegraphics[width=.95\textwidth]{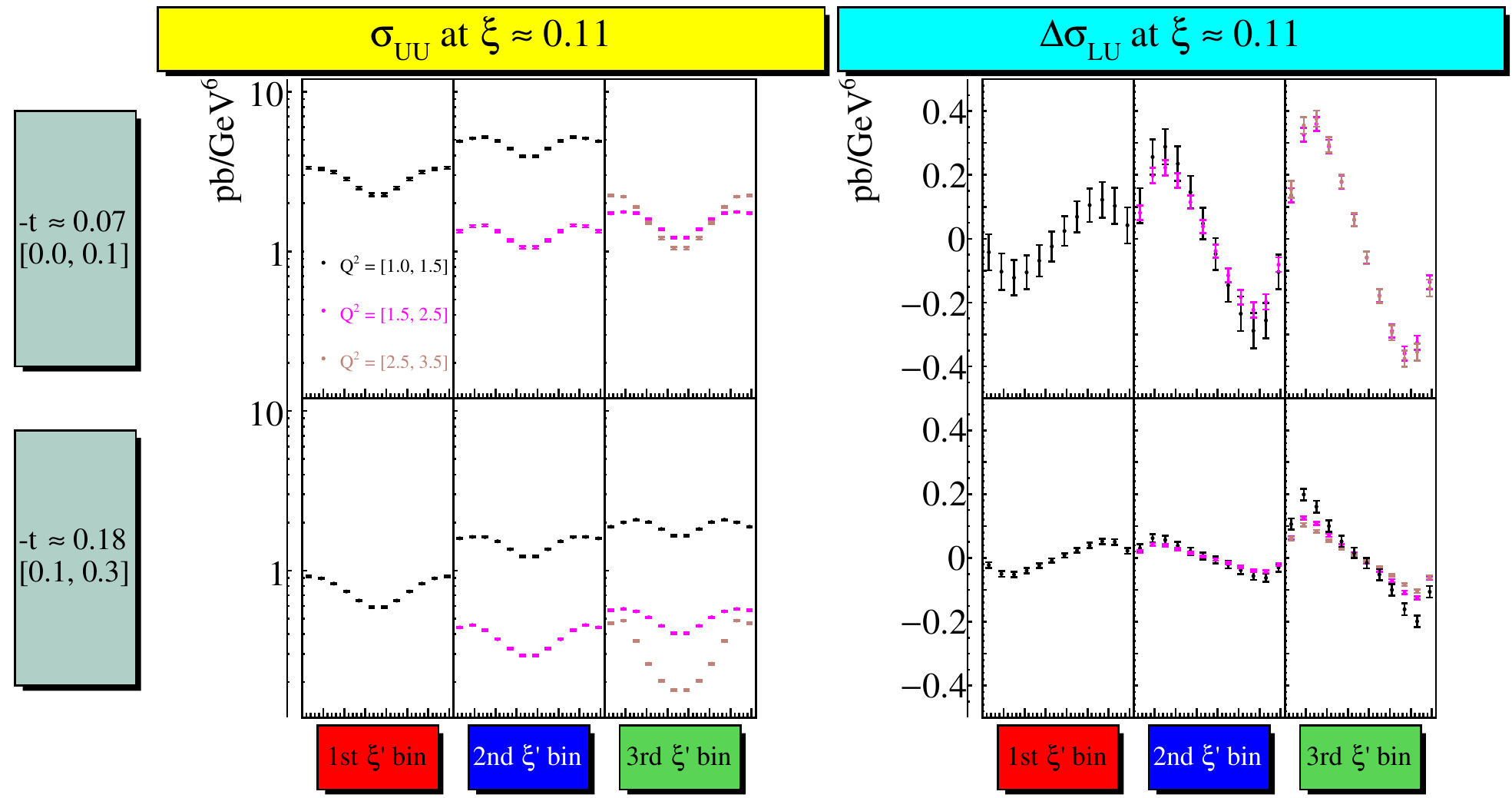} 
\caption{Unpolarized cross section $\sigma_\text{UU}$ (left half) and beam spin cross section difference $\Delta\sigma_\text{LU}$ (right half) as a function of $\phi$ at $\xi\approx0.11$. The panels of the top row and bottom row show them for $-t=[0,0.1]$ and $[0.1,0.3~$(GeV/$c^2$)$^2$], respectively. The columns stand for the three $\xi'$ bins in Fig.~\ref{fig1}. The black, magenta and brown dots correspond to $Q^2=[1,1.5]$, $[1.5,2.5]$ and $[2.5,3.5~$(GeV$/c^2$)$^2$], respectively.}
\label{fig2}
\end{figure}

The experimental observables have been evaluated by a DDVCS event generator developed with the VGG model\cite{VGG1,VGG2} at leading twist. Kinematic cuts to ensure the applicability of the GPD formalism have been applied\cite{ZHAO}. We have adopted a 5-dimensional grid of bins in $\xi'$, $\xi$, $Q^2$, $t$ and $\phi$ covering the whole kinematic phase space of interest. The right panel of Fig.~\ref{fig1} depicts the grid of bins in ($\xi',\xi$), where three bins indicated by 1, 2 and 3 have approximately the same average $\xi=0.11$ and different average $\xi'$. In the following, the experiment projections and the extraction of CFFs are discussed with respect to the three $\xi'$ bins.
 
The projections have been performed in the ideal situation that all the particles of the final state can be detected with 100\% efficiency. The count number has been calculated for a luminosity $\mathrsfso{L} = 10^{36} \text{cm}^{-2}\cdot\text{s}^{-1}$ considering 50 days running time. In this work, only unpolarized cross section $\sigma_\text{UU}$ and beam spin cross section difference $\Delta\sigma_\text{LU}$ are discussed. Figure \ref{fig2} shows them with statistic errors as a function of $\phi$ for the three $\xi'$ bins, two bins in $-t~[0,0.1,0.3~($GeV$/c^2)^2$] and three bins in $Q^2~[1,1.5,2.5,3.5~($GeV$/c^2)^2$]. It indicates that it is possible to obtain DDVCS experimental observables with good precision.

\begin{figure}[b]
\centering
\includegraphics[width=1\textwidth]{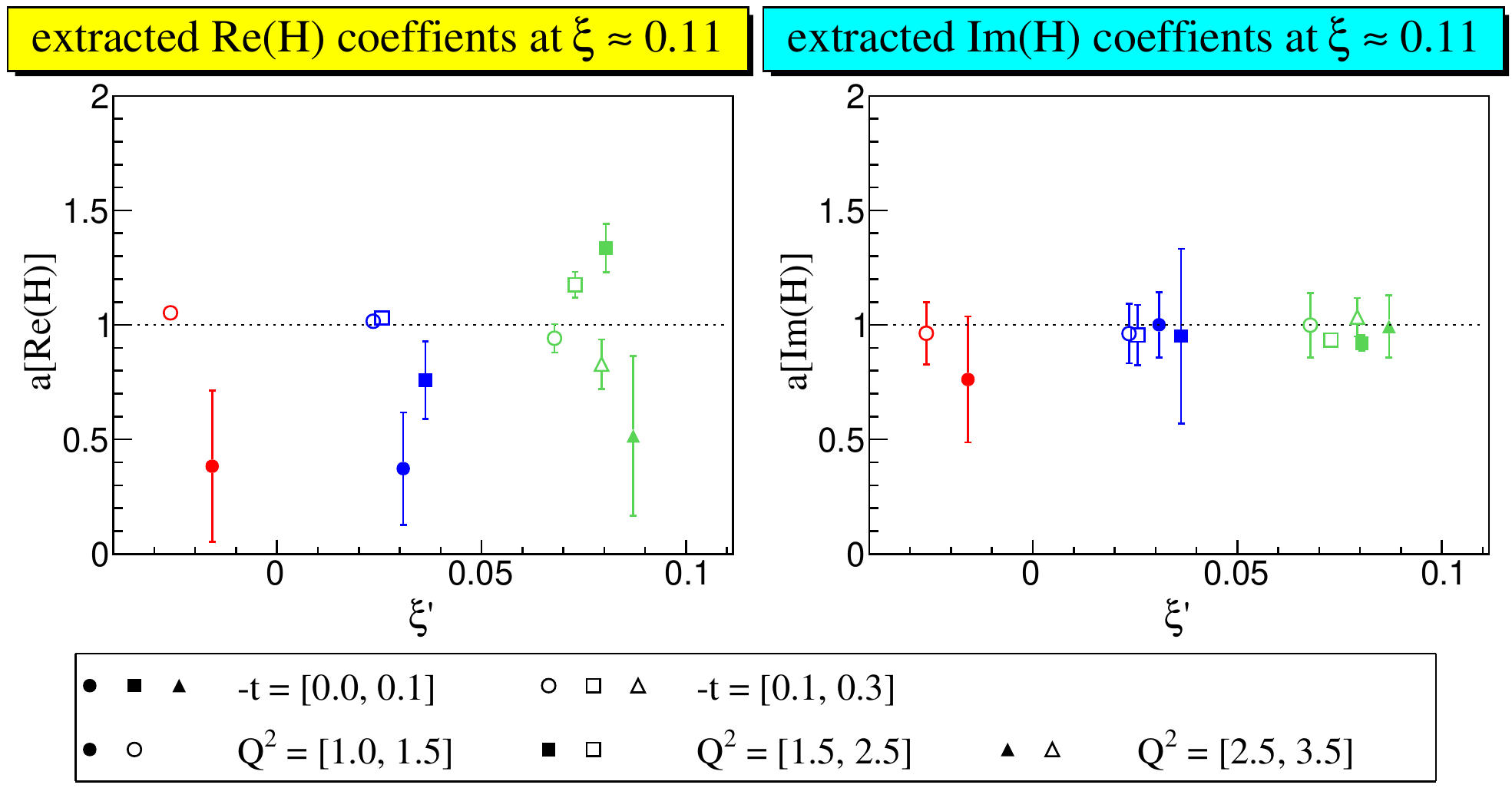} 
\caption{The extracted CFF coefficients with respect to the observables in Fig.~\ref{fig2}. The left panel is the real part of CFF $\mathcal{H}$ while the right one is the imaginary part.}
\label{fig3}
\end{figure}

\section{CFF extraction}
\label{sec3}

The fitting program is inspired from the DVCS one's\cite{Fit1}, corresponding to a quasi-model-independent way to extract CFFs. It consists in taking the eight CFFs as free parameters, and knowing the well-established BH and DDVCS leading-twist amplitudes, to fit, at a fixed kinematics, simultaneously the $\phi$-distributions of several experimental observables. If only $\sigma_\text{UU}$ and $\Delta\sigma_\text{LU}$ are available, only CFF $\mathcal{H}$ can be well extracted, with $\sigma_\text{UU}$ being particularly sensitive to the real part and $\Delta\sigma_\text{LU}$ dominated by the imaginary part. Figure \ref{fig3} shows the extracted CFF coefficient, defined as the fitted CFF value normalized by the one of the event generator. The dashed line indicates the coefficient being 1, which leads to the fact that the CFF is perfectly recovered. The imaginary part looks obviously well recovered, which means that we are able to extract the imaginary part of $\mathcal{H}$. With respect to the real part, there is a relatively large discrepancy between the fitted and the generated CFF, due to the complex component of $\sigma_\text{UU}$ from different contributions. In order to perform a better extraction of the real part, the beam charge cross section is required since it involves only the interference contribution, which is linear in the CFFs \cite{DDVCS3,ZHAO}.

\begin{figure}[t]
\centering
\includegraphics[width=1\textwidth]{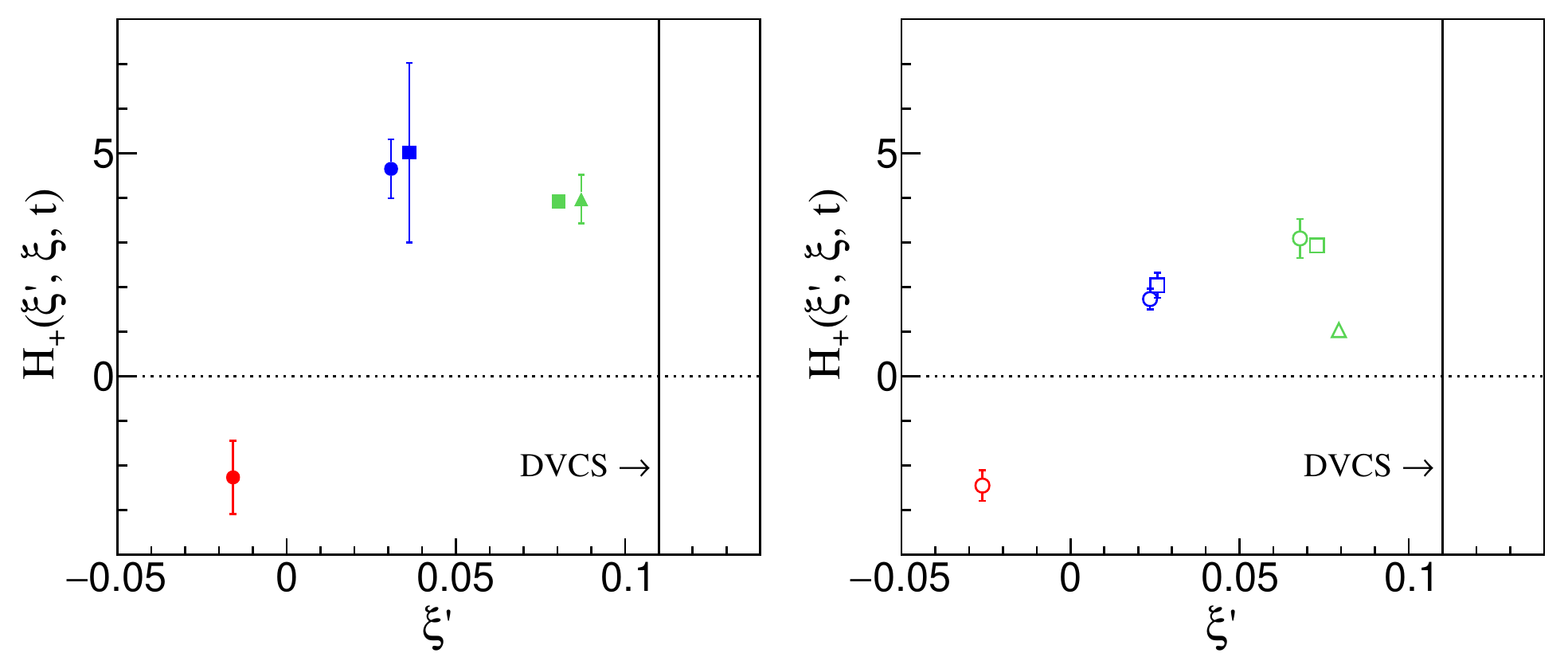} 
\caption{\label{fig_singlet}The proton singlet GPD $H_+$. The left panel stands for $-t=[0,0.1~$(GeV/$c^2$)$^2]$ and the right one for $[0.1,0.3~$(GeV/$c^2$)$^2]$. The convention
for points is the same as in Fig.~\ref{fig3}.}
\end{figure}

Eventually, the proton singlet GPD $H_+$ can be extrapolated from the imaginary part of CFF $\mathcal{H}$. Figure \ref{fig_singlet} shows $H_+(\xi',\xi,t)$ as a function of $\xi'$ at fixed $\xi$ for two bins in $t$ and three bins in $Q^2$. The vertical solid line stands for the location accessed by DVCS and the colored points extracted from the DDVCS pseudo-data give the decoupled singlet GPD value.

\section{Conclusion}
\label{sec4}

Our study shows a high degree of feasibility of a DDVCS experiment at a relative challenging luminosity with exclusive final states completely detected. We also have the capability to access decoupled singlet GPD. The fitting program is still confronted with a severely underconstrained problem and running time difficulty, and further work is still ongoing.

\end{document}